\documentclass[lettersize,journal]{IEEEtran}
\usepackage{amsmath,amsfonts}
\usepackage{algorithmic}
\usepackage{algorithm}
\usepackage{array}
\usepackage[caption=false,font=normalsize,labelfont=sf,textfont=sf]{subfig}
\usepackage{textcomp}
\usepackage{stfloats}
\usepackage{url}
\usepackage{verbatim}
\usepackage{graphicx}
\usepackage{cite}
\begin{document}

\title{A Low-Cost Reliable Racetrack Cache Based on Data Compression}

\author{Elham Cheshmikhani, Fateme Shokouhinia, and Hamed Farbeh,~\IEEEmembership{Member,~IEEE,}
\thanks{E. Cheshmikhani is with the Department of Computer Science and Engineering, Shahid Beheshti University, Tehran, Iran. E-mail: e\_cheshmikhani@sbu.ac.ir\\
F. Shokouhinia is with the Department of Computing Science, Simon Farser University, BC, Canada. Email: fateme\_shokouhinia@sfu.ca\protect\\
H. Farbeh is with the Department of Computer Engineering, Amirkabir University of Technology, Tehran, Iran. 	E-mail: farbeh@aut.ac.ir}
\thanks{Manuscript received October 25, 2023.}}

\markboth{IEEE Transactions on Circuits and Systems—II: Express Briefs (TCAS-II),~Vol.~14, No.~8, October~2023}%
{Shell \MakeLowercase{\textit{et al.}}: A Sample Article Using IEEEtran.cls for IEEE Journals}


\maketitle

\begin{abstract} 
SRAM-based cache memory faces several scalability limitations in deep nanoscale technologies, e.g., high leakage current, low cell stability, and low density. Emerging Non-Volatile Memory (NVM) technologies have received lots of attention in recent years, where \textit{Racetrack~Memory} (RTM) is among the most promising ones. 
RTM has the highest density among all NVMs and its access performance is comparable to SRAM technology. 
Therefore, RTM is a suitable alternative for SRAM in the Last-Level Caches (LLCs). 
Despite all its benefits, RTM confronts different reliability challenges due to the stochastic behavior of its storage element and highly error-prone data shifting, leading to a high probability of multiple-bit errors. 
Conventional Error-Correcting Codes (ECCs) are either incapable of tolerating multiple-bit errors or require a large amount of extra storage for check bits. 
This paper proposes taking advantage of value locality for compressing data blocks and freeing up a large fraction of cache blocks for storing data redundancy of strong ECCs. 
Utilizing the proposed scheme, a large majority of cache blocks are protected by strong ECCs to tolerate multiple-bit errors without any storage overhead. 
The evaluation using gem5 full-system simulator demonstrates that the proposed scheme enhances the \textit{mean-time-to-failure} of the cache by an average of 11.3x with less than 1\% hardware and performance overhead.
\end{abstract}

\vspace{-10pt}
\section{Introduction}
In recent years, with technology's downscaling trend and high-performance AI applications, there has been a surge in research and development efforts aimed at improving the performance and efficiency of computing systems.
Cache memories as a vital component of modern computing architectures play a vital role in bridging the gap between fast but small processor registers and slow but large main memory.
Traditionally, cache designs have focused on utilizing $Static~Random$-$Access$ $Memory$ (SRAM) cells to store frequently accessed data.
However, due to the limitations of scaling, increasing power demands, and their volatility, emerging \textit{Non-Volatile Memories} (NVMs) including \textit{Racetrack Memory} (RTM), also known as \textit{Domain-Wall Memory} (DWM), have gained recognition as a potential replacement for SRAM in cache designs.

RTM caches leverage the physics of magnetic domains to store and retrieve data, offering several potential advantages over conventional SRAM caches.
These advantages include extremely higher storage density, lower power consumption, non-volatility, and reduced susceptibility to radiation-induced errors, making them particularly appealing for both low-power and high-performance computing systems \cite{hameed2022blendcache}.  However, despite these promising benefits, RTM faces several reliability challenges including shift errors and domain-wall tilting \cite{boulle2013domain}.

The core principle behind domain RTM caches is the utilization of nanoscale magnetic domains, called $domain~walls$, to represent and store binary data \cite{khan2023downshift}.
The movement of these domain walls along a magnetic nanowire creates resistance variations that can be sensed and interpreted as digital states \cite{kline2019premsim}.
Each domain contains a data bit accessed via access ports.
As there are few access ports for several domains, shifting domains is required to be aligned with the access ports.

The wrong number of shifts, named \textit{shift error}, results in accessing the wrong data in RTM.
\textit{Tilting error} refers to the misalignment of cells in the domain \cite{boulle2013domain, li2023ultralow}.
In addition to RTM-specific errors, all error sources in Spin-Transfer Torque MRAM (STT-MRAM), as the RTM predecessor, including \textit{retention~failure}, \textit{write~failure}, and \textit{read~disturbance} are also probable to occur in RTMs, which make RTM as the most error-prone memory technology \cite{cheshmikhani20213rset}.


Several studies addressed various sources of errors in RTM and tried to mitigate their occurrence rate \cite{archer2020foosball, zhang2015hi,ollivier2019leveraging, kline2019premsim, khan2023downshift, wang2016np, vahid2017correcting, chee2019codes1}.
However, to make it applicable, an error protection/recovery mechanism is necessary besides the efforts for error rate reduction.
Employing \textit{Error}-\textit{Correcting~Codes} (ECCs) is the most widespread approach for LLC safeguarding.
However, the conventional \textit{Single}-\textit{Error~Correction~and~Double}-\textit{Error~Detection} (SEC-DED) 
codes do not offer enough protection against multiple-bit errors that are likely to occur in RTM caches.
On the other hand, more robust ECCs that can handle higher levels of error correction require a significant amount of storage, making them impractical for on-chip caches.

This paper presents a solution for protecting RTM cache blocks using ECCs that can be arbitrarily strong, without the need for additional hardware storage for the check bits.
To achieve this, we take advantage of the value locality present in the block's content to compress the data and store it in a reduced number of cells.
Furthermore, the unused portions of the blocks are utilized for storing the redundancy needed for the strong ECC.
By offering robust protection for the compressible data that makes up the majority of cache blocks, we significantly improve the reliability of the RTM cache.

The proposed scheme is simulated using the gem5 full-system simulator \cite{lowe2020gem5} and SPEC CPU2017 benchmark suite \cite{bucek2018spec}.
The evaluation based on a quad-core processor shows that the proposed scheme prolongs the mean-time-to-failure (MTTF) of RTM-based LLC by 11.3x compared to its SEC-DED-protected counterpart.
This significant reliability enhancement is achieved while incurring less than 1\% overhead.



\vspace{-5pt}
\section{RTM Basics and Challenges}
\subsection{RTM Cell Architecture}
\begin{figure}[t]\vspace{-10pt}
	\captionsetup{font=footnotesize}
	\centering
	\includegraphics[width=0.73\linewidth]{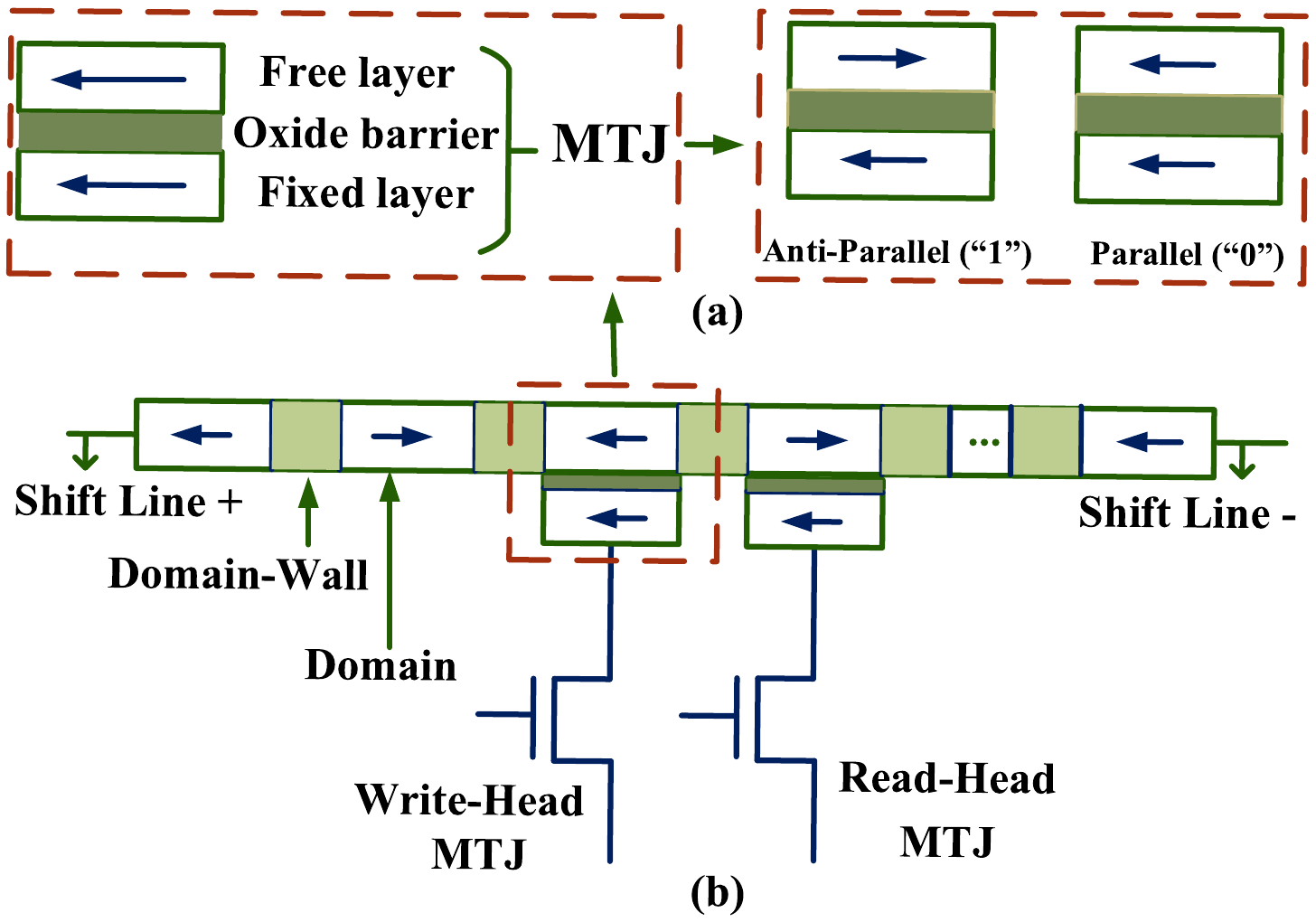}\vspace{-5pt}
	\caption{a) MTJ structure and b) A simple RTM nanowire with several domains.}\vspace{-13pt}
	\label{fig1}
\end{figure}

RTM has an array structure consisting of nanowires placed on a silicon chip \cite{blasing2020magnetic}. These nanowires are made up of domains with walls between them. Each domain has a magnetization state to determine the data stored in it (0 or 1). RTM can be extremely dense since each device can contain multiple domains and hence, multiple data. The read/write operations are done with the help of access transistors (access ports). For aligning an access port to a domain, RTM needs shift operation requiring applying shift current \cite{boulle2013domain}.
A simple nanowire of RTM with several domains and two access ports is demonstrated in Fig. \ref{fig1}-b \cite{li2023ultralow}, which its MTJ structure is magnified in Fig. \ref{fig1}-a.
MTJ consists of three layers: \textit{free, oxide barrier}, and \textit{fixed layers}. The same magnetization directions of the free and fixed layers represent “0” and vice versa. Besides, each domain of RTM plays the role of the free layer when connected to an access transistor. In the upcoming subsections, we will describe read, write, and shift operations in RTM.

\subsubsection{Read and Write Operation}
Read and write operations are similar between RTM and STT-MRAM since they both use the MTJ structure as their storage element.
MTJ’s detailed structure is shown in Fig. \ref{fig1}-a.
The read process contains three main steps: (a) turning on Word Line (WL) in the access transistor, (b) applying read current ($I\textsubscript{read}$) through MTJ, and (c) comparing the voltage between Source Line (SL) and Bit Line (BL) with a reference voltage \cite{roxy2020novel}.
The stored data is interpreted as “0” (“1”) if this voltage is lower (higher) than the reference voltage\cite{cheshmikhani20213rset}.

The write process results in flipping the magnetic state of a domain, which is the free layer in the MTJ.
When a current is applied to BL (SL), it changes the free layer’s magnetic orientation.
This current can be different from (the same as) the current flow, so the value “1” (“0”) is written into the cell \cite{cheshmikhani20213rset}.
The writing process in the new version of RTM performed based on shifting to improve the write performance \cite{venkatesan2015cache}.

\subsubsection{Shift Operation}
Domains must use shift operations to align to the access ports in RTM. To this aim, a shift current with a specific magnitude is applied through the shift transistors placed at each end of a nanowire. This current causes a phenomenon called \textit{domain-wall motion} that results in the shifting of domains \cite{kumar2022domain}.
Shift operation causes various overheads and affects the performance of RTM. Extra delay and energy will be needed for each access depending on the number of shifts it takes. Moreover, RTM suffers from area overhead as each track requires 50\% vacancy to prevent data loss during domain shifting \cite{khan2023downshift}.
Domain shifting also poses reliability issues elaborated in the following section.

\vspace{-10pt}
\subsection{Reliability Challenges and Solutions}
Several factors impair the reliability of RTM, such as (a) domain shifting and (b) MTJ structure (where the domains act as the free layer) \cite{kline2019premsim}. We describe these factors in depth in the subsequent subsections.
\subsubsection{Position Errors}
\begin{figure}[t]\vspace{-10pt}
	\captionsetup{font=footnotesize}
	\centering
	\includegraphics[width=0.87\linewidth]{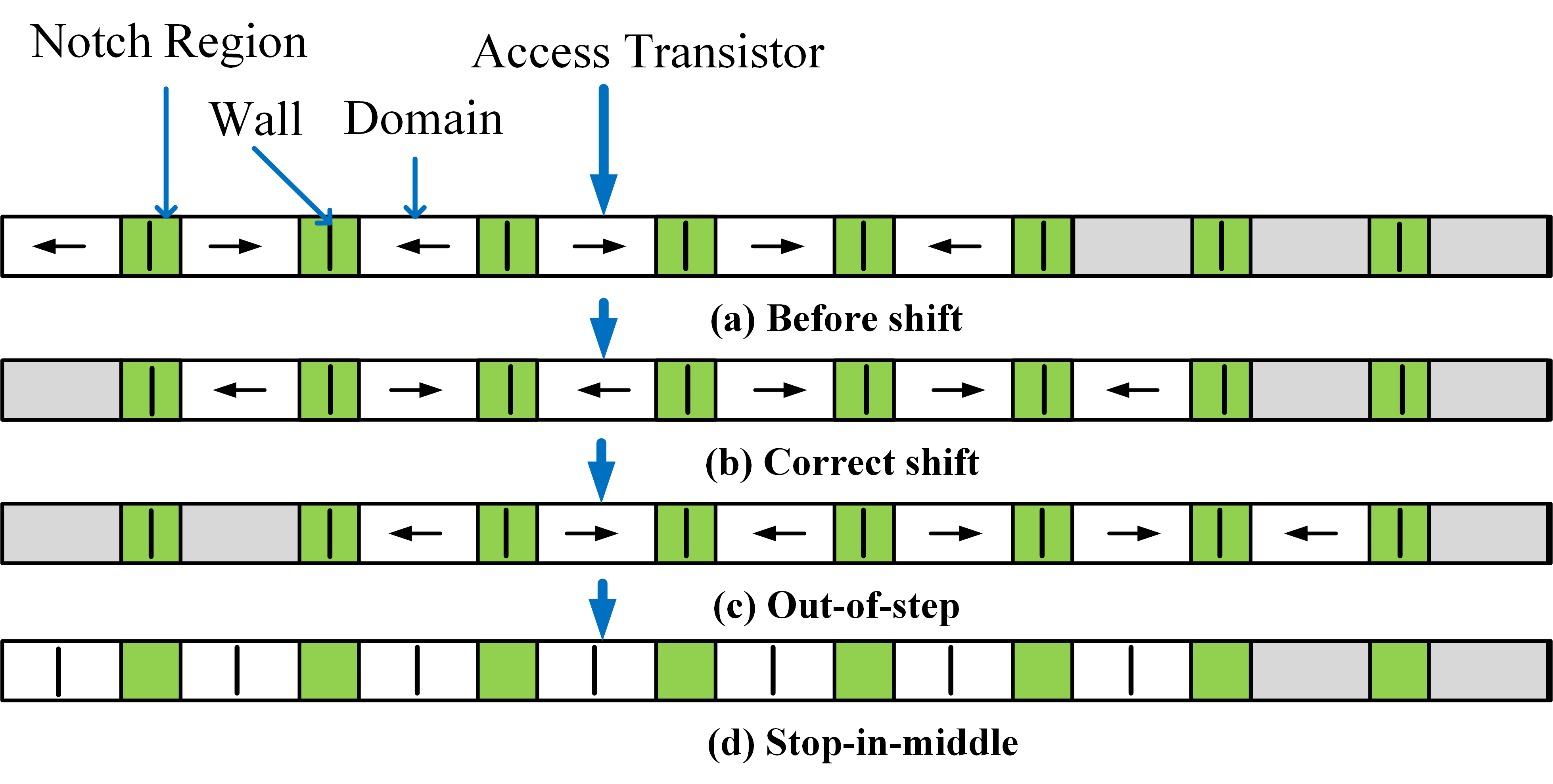}\vspace{-8pt}
	\caption{RTM cells, correct shift operation, and possible shift errors \cite{zhang2015hi}.}\vspace{-13pt}
	\label{fig2}
\end{figure}

Domain shifting is required to match each domain with a suitable access transistor. 
This operation entails energy and latency costs as well as reliability threats. 
A correct shift operation depends on the alignment of the desired domain to the access port and the proper connection of notch regions and walls.
Therefore, domain shifting can lead to two errors: (a) shifting more or less than required and (b) misalignment of the domain and not properly attachment of notch regions and walls.
Fig. \ref{fig2} illustrates these errors and contrasts them with the correct shifting state.
These errors are termed \textit{Position Errors}, with the former being \textit{out-of-step shifting} as shown in Fig. \ref{fig2}-c and the latter being \textit{stop-in-middle} as shown in Fig. \ref{fig2}-d \cite{zhang2015hi, kline2019premsim, khan2023downshift}.

\subsubsection{MTJ Structure Errors}
The MTJ structure in RTM is similar to STT-MRAM.
This resemblance implies that they have common reliability challenges and errors arising from MTJ, although this aspect has been overlooked in prior studies.
The errors in STT-MRAM are classified into three categories: (a) \textit{Write Failure}, (b) \textit{Read Disturbance}, and (c) \textit{Retention Failure} \cite{cheshmikhani2019system, hadizadeh2020stair}. Write failure occurs when the write operation fails to switch the cell content correctly \cite{cheshmikhani2019robin}. Read disturbance occurs when the read current inadvertently alters the cell content \cite{hadizadeh2021copa, cheshmikhani2019enhancing}. Retention failure occurs when a cell changes its content randomly after being idle for some time (idle means not read and not written) \cite{hadizadeh2021copa}.

In addition to these errors shared with STT-MRAM, RTM may encounter other potential errors, such as domain-wall tilting. 
Domain-wall tilting errors can arise from the interaction of some magnetic effects during domain shifting \cite{boulle2013domain}.
These errors depend on the physical characteristics of RTM memories.
\subsubsection{Related Work}
As mentioned, position errors and MTJ structure errors are the main reliability threats.
In this subsection, we explain the studies that addressed errors due to the MTJ structure in STT-MRAM and position errors in RTM. 
The Sub-Threshold Shifting method (STS) is proposed as a solution for the stop-in-middle error 
\cite{zhang2015hi}. 
STS employs a two-step shift current to ensure the completion of the shift operation.
This approach increases the probability of other errors such as out-of-step shifting.
Position-Errors Correction Code (P-ECC) was introduced to detect and correct two-step shift errors \cite{zhang2015hi}. This method employs additional domains at the end of each track to store P-ECC bits. It also requires extra read ports for added parts, which incur high overhead.

P-ECC has been enhanced in \cite{wang2016np} to correct one bit-flip error besides shift errors. Additional access ports and a large number of extra bits entail significant performance and area costs in this method. The overhead grows exponentially when more than one error correction is needed.
In \cite{vahid2017correcting}, these overheads are removed by separating error detection from correction. Using effective coding schemes such as \cite{chee2019codes1}, correcting d-step errors only requires d+1 additional access ports and one extra domain.

Derived Error-Correction Coding (D-ECC) was proposed for protecting RTM from alignment faults \cite{ollivier2019leveraging}. This method employs the Traverse Read (TR) approach, which counts the number of “1”s in an RTM track, and then uses this number for an ECC called Traverse-read ECC (TECC). By extending TECC, DECC is obtained with lower dynamic energy and area costs than P-ECC. Moreover, DECC uses STT-MRAM for storing ECC bits, which eliminates area overhead.
In \cite{archer2020foosball}, authors incorporate Hamming codes into GreenFlag to enable bit-flip error detection and correction. Consequently, SEC-DED is achieved for both shift errors and bit-flip errors. This enhanced method is termed Foosball Coding. This is complex coding and entails high area and energy costs.

ECCs are employed in various ways to overcome MTJ structure errors.
In \cite{cheshmikhani2019enhancing}, authors reduced read disturbance by applying ECC for all the blocks in a cache.
ROBIN attempted to use all bits of all ways and all blocks to generate ECCs to mitigate write failure error \cite{cheshmikhani2019robin}.

In conclusion, RTM-based solutions entail significant area and energy costs and these studies neglect errors related to MTJ structure shared between STT-MRAM and RTM as well as permanent faults due to limited endurance.
Moreover, STT-MRAM-based methods require extensive modifications to be compatible with RTMs.

\section{Motivation and Proposed Method}
Racetrack memory faces several sources of errors due to its unique structure. Position or shift errors are among the main types of errors.
Additionally, since RTM is MRAM-based, it inherits error types from STT-MRAM.
Previous studies on improving the reliability of RTM have focused on reducing the rate of certain errors or tolerating single-bit errors. However, the diverse sources of errors and the high susceptibility of RTM necessitate the implementation of fault-tolerant mechanisms in addition to error reduction schemes. Furthermore, the high rate of errors from various sources often results in multiple-bit errors, which require stronger protection.
The conventional method for achieving fault-tolerance in on-chip caches is through the use of ECCs. However, the widely used \textit{Single}-\textit{Error~Correction~and~Double}-\textit{Error~Detection} (SEC-DED) coding scheme is unable to handle multiple-bit errors, and more robust ECCs are not feasible in caches due to the significant amount of additional redundancy they require. 

To address these issues, this paper proposes a highly reliable RTM cache that employs strong ECCs without the need for extra storage overhead. Our proposed scheme utilizes data compression to store incoming data blocks in smaller spaces, which frees up a substantial number of cache lines for storing redundant ECC bits. By opportunistically utilizing data bits in compressed blocks for check-bit storage, we can utilize strong ECCs and effectively correct multiple-bit errors in data blocks. The details of our proposed scheme will be discussed in the following section.



\begin{table}[t]
				\centering
				\caption{Comparison of various data compression schemes.}\vspace{-5pt}
				\includegraphics[width=1\linewidth]{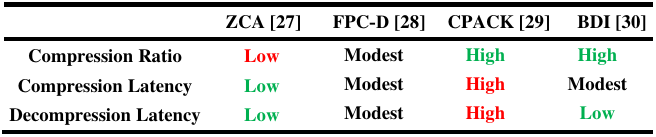}
				\label{table1}\vspace{-20pt}
\end{table}
\vspace{-8pt}
\subsection{Data Compression for ECC storage}\vspace{-2pt}
Compressing data to fit into smaller storage spaces has been a topic of interest in various applications for several decades. In recent years, several studies have explored the use of data compression in on-chip caches. The objective of these studies has been to increase the effective cache capacity to enhance performance and/or reduce energy consumption. Among them, only \cite{xu2015multilane} targeted RTM-based caches. Our proposed scheme is the first to utilize data compression in RTM caches for reliability enhancement.


The primary criterion for our proposed scheme is to employ an efficient data compression algorithm. Low compression/decompression latency and high compression ratio are required for an efficient algorithm. Additionally, our scheme requires a binary decision on whether to compress incoming data. In cases where the reduction in the number of data bits does not meet the required number of check bits, compression is not applied. In order to determine the most suitable compression algorithm, we have examined state-of-the-art algorithms.


These algorithms can be classified into two categories: \textit{cache compaction} and \textit{cache compression} algorithms.
While the latter focus on representing data using fewer bits and freeing up a portion of the cache line \cite{dusser2009zero, alameldeen2018opportunistic, chen2009c, pekhimenko2012base}, the former tries to exploit the freed parts to accommodate multiple compressed data blocks into a single cache line \cite{ghasemazar20202dcc, park2021bcd, ghasemazar2020thesaurus}. 
The proposed scheme is interested in cache compression algorithms that can make use of the freed cache line space for ECC storage.


Table \ref{table1} compares different compression algorithms in terms of their overhead and capability to detect data patterns \cite{dusser2009zero, alameldeen2018opportunistic, chen2009c, pekhimenko2012base}. It is evident that BDI is the best choice for our goal due to its higher compression ratio by detecting more data patterns and lower decompression latency, despite its non-minimal compression latency. It is noteworthy that compression latency is not of utmost importance as this operation is not on the critical path of LLC accesses and can be compromised in favor of compression rate and decompression latency.

\vspace{-5pt}
\subsection{Architectural Details} 
\vspace{-2pt}
Equipping LLC with a data compressor frees up enough space to store strong ECCs alongside the compressed data blocks. The proposed scheme offers two configurations for the cache structure: 1) no additional storage is embedded for check bits, and only compressed blocks receive protection from our strong ECC, and 2) all cache blocks are protected by conventional SEC-DED code stored in low-cost extra storage, while the compressed blocks are protected by our strong ECC. The second configuration is preferred because leaving even a small fraction of cache blocks unprotected in error-prone RTM-based caches can significantly reduce the reliability of the system. We assume the use of \textit{Triple Error Correction-Quad Error Detection} (TEC-QED) for strong ECC protection, although any other ECC  is also applicable as long as the number of check bits resides in the compressed part.


\begin{figure}[t]
	\captionsetup{font=footnotesize}
	\centering
	\includegraphics[width=0.73\linewidth]{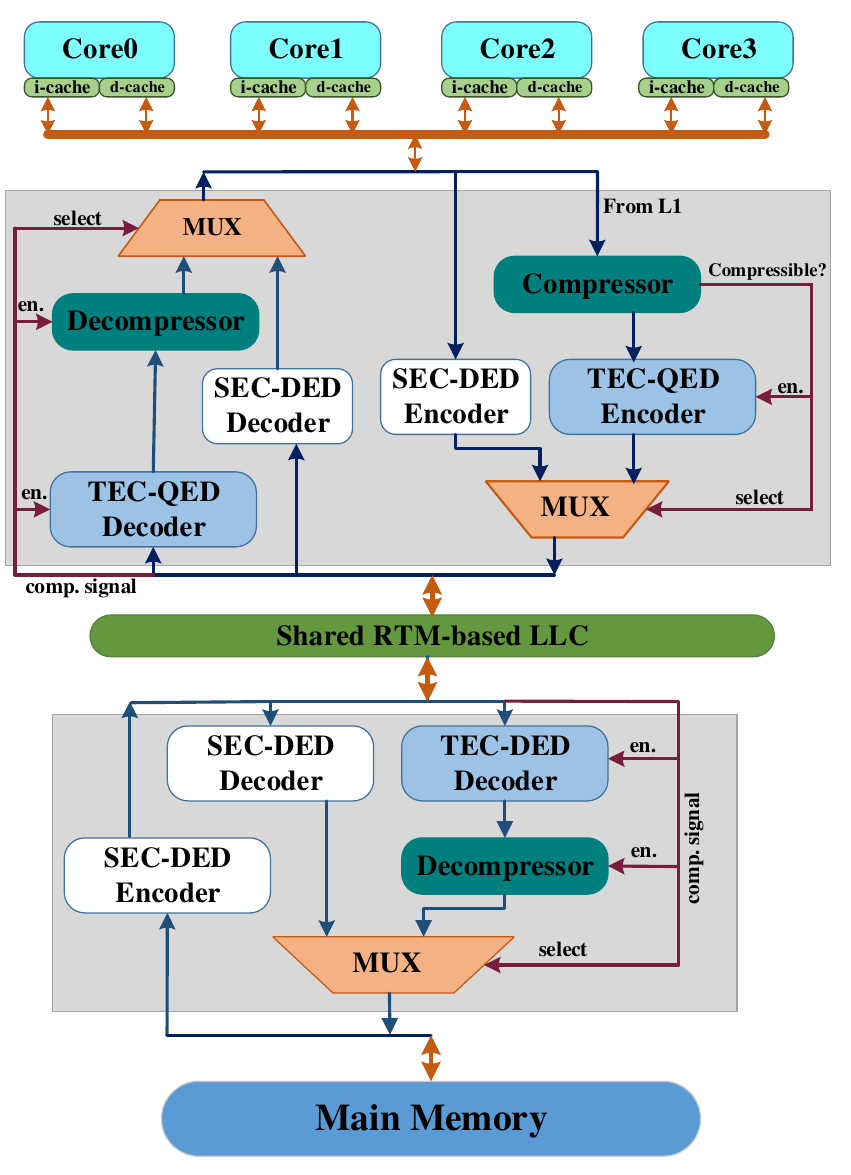}\vspace{-10pt}
	\caption{Proposed reliable RTM cache architecture.}\vspace{-15pt}
	\label{fig3}
\end{figure}

Despite the high compression ratio of BDI \cite{pekhimenko2012base}, it cannot guarantee to compress all blocks. Hence, a fraction of blocks remains vulnerable to multiple-bit errors. However, data blocks in the cache are classified into $dirty$ or $clean$ blocks. On detecting an error in clean blocks, the data can easily be recovered by re-fetching it from a lower memory level without the need for in-site data correction. To maximize the error recovery rate and minimize the overheads of data compression, the proposed scheme only applies compression and strong ECC on dirty blocks and the clean blocks are recovered by invalidating them and re-fetching them from the main memory.

Fig. \ref{fig3} illustrates the architectural details of the proposed scheme. Data is written into LCC from two directions, i.e., from main memory on cache misses and from write-backs of higher cache levels. 
Blocks that arrive from the main memory do not require compression.
Therefore, only the blocks written back from the higher cache level undergo compression checking. 
If a block can be compressed, it is sent to the strong-ECC encoder unit, and both check bits and compressed block are stored in a cache line. Otherwise, the uncompressed data is protected by the generated default SEC-DED code. 


When reading data blocks, there are two scenarios: reading requests from higher cache levels and evicting a dirty block. In the former case, if it is a clean or uncompressed dirty block, the block and its corresponding SEC-DED code are sent to the SEC-DED decoder for error checking. Otherwise, error checking is performed by the strong-ECC decoder unit, and the decompressor unit generates the original data.



\vspace{-5pt}
\section{Simulation Setup and Results}\vspace{-3pt}
\subsection{Simulation Setup}\vspace{-3pt}

The proposed scheme is implemented in the gem5 full-system simulator \cite{lowe2020gem5} modelling a quad-core ARM processor equipped with a per-core SRAM-based 32-KB 4-way L1 I/D cache and a shared RTM-based 2MB 16-way LLC cache. A set of memory-intensive programs from SPEC CPU2017 \cite{bucek2018spec} are used in different combinations as the workload. The simulations are performed for one Billion instructions after skipping the first 100 Million instructions as the warm-up.

The proposed scheme is compared with an RTM cache protected by SEC-DED code as the baseline. The encoder/decoder latency of SEC-DED and TEC-QED code is one clock cycle and three clock cycles, respectively. The compression/decompression latency of BDI is 2/1 clock(s).















%

%










\begin{figure}[t]
\captionsetup{font=footnotesize}
\centering
\includegraphics[width=0.9\linewidth]{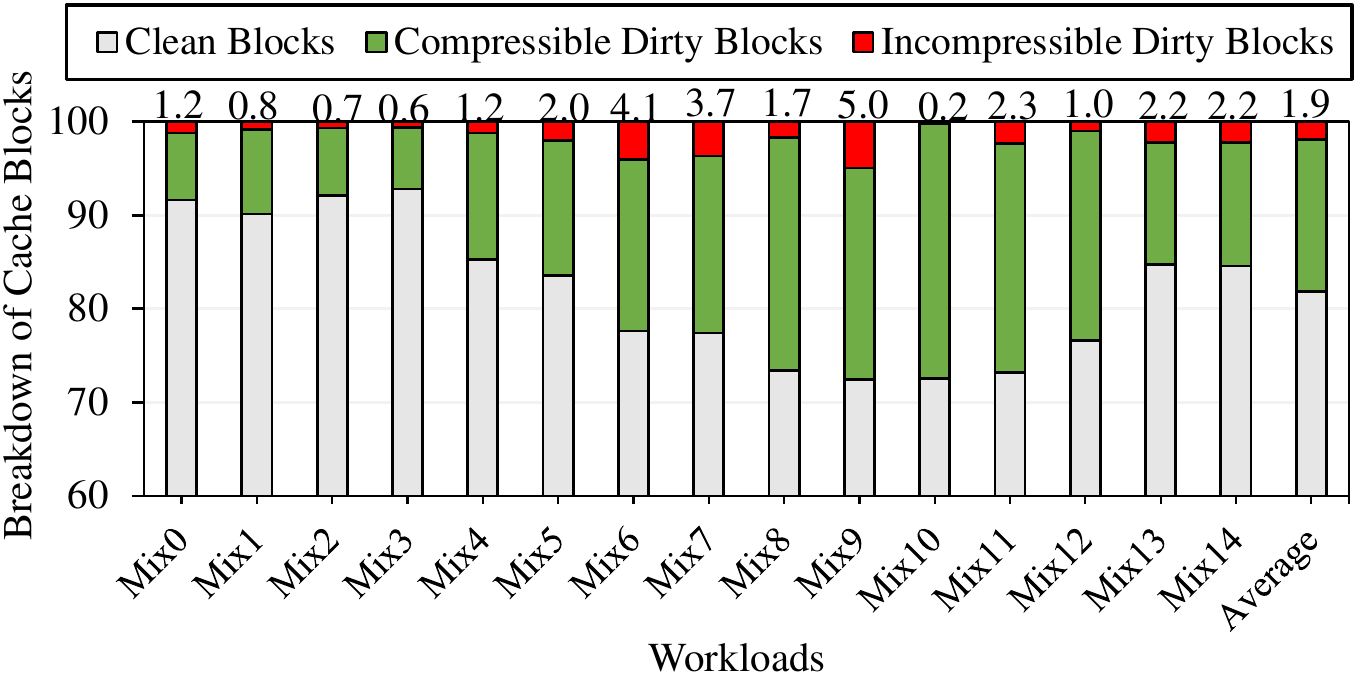}\vspace{-10pt}
\caption{Contribution of clean and dirty (compressible/incompressible) blocks.}\vspace{-15pt}
\label{fig-break}
\end{figure}

\vspace{-8pt}
\subsection{Results}\vspace{-3pt}
Fig. \ref{fig-break} demonstrates the fraction of cache blocks that are vulnerable to uncorrectable multiple-bit errors in the proposed scheme and SEC-DED.
While both schemes can withstand multiple-bit errors in clean blocks, the SEC-DED cache provides no multiple-bit error recovery for dirty blocks and the proposed scheme tolerates the errors in compressed blocks.
According to the results, an average of 18.1\% of the blocks are vulnerable in the baseline and the proposed scheme reduces this value to 1.9\%. Hence, our scheme decreases the fraction of vulnerable blocks to less than one-ninth.

\begin{figure}[t]
\captionsetup{font=footnotesize}
\centering
\includegraphics[width=0.99\linewidth]{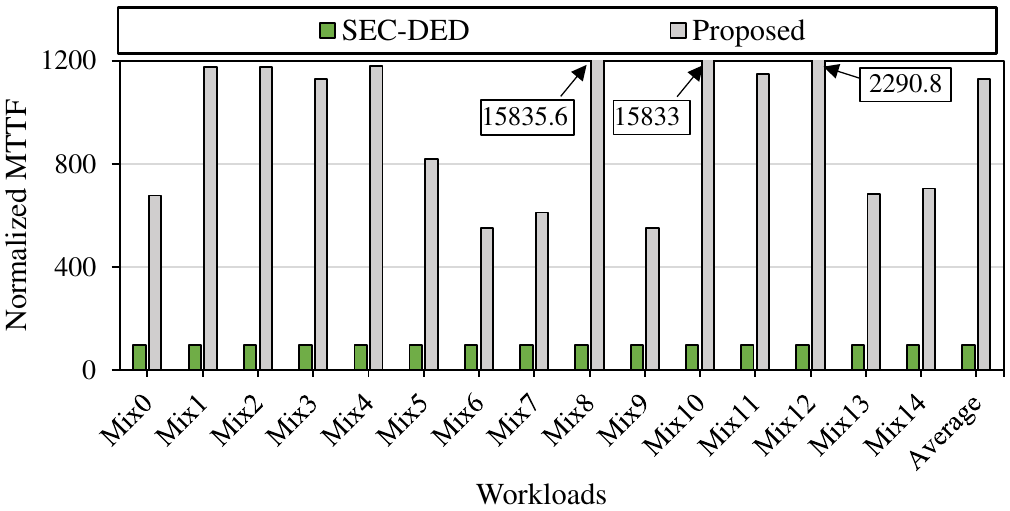}\vspace{-10pt}
\caption{MTTF of the proposed RTM cache normalized to the baseline cache.}\vspace{-5pt}
\label{fig-mttf}
\end{figure}

As a reliability metric, Fig. \ref{fig-mttf} illustrates the mean-time-to-failure (MTTF) of the cache for the proposed scheme, normalized to the baseline.
On average, MTTF is prolonged by 11.3x, highlighting a significant reliability enhancement.
The MTTF increase reaches up to 158.3x in $mix8$ workload and to 5.5x in the worst-case for $mix6$ workload.

\begin{figure}[t]
\captionsetup{font=footnotesize}
\centering
\includegraphics[width=1\linewidth]{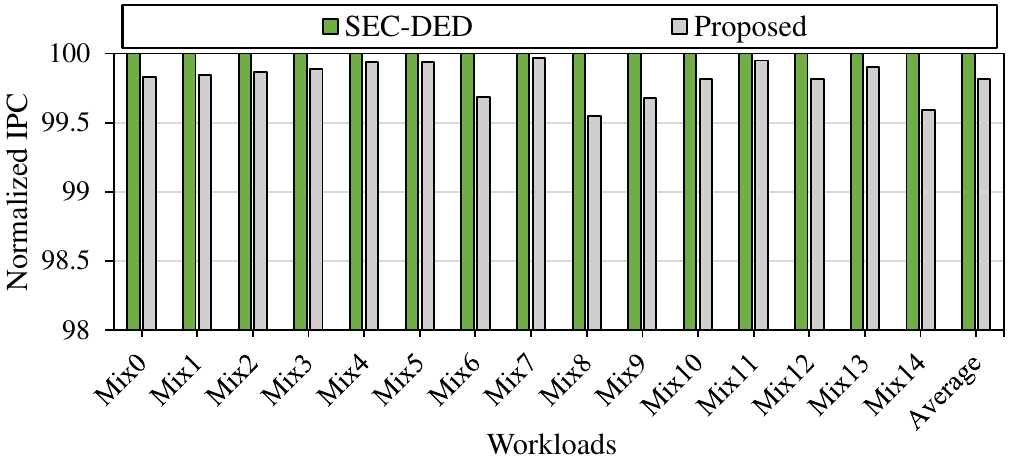}\vspace{-8pt}
\caption{IPC of the proposed RTM cache normalized to the baseline cache.}\vspace{-15pt}
\label{fig-ipc}
\end{figure}

The proposed scheme may introduce performance overhead by increasing the cache access time for certain requests.
As explained in Section 4, the (de)compressor and strong-ECC encoder/decoder units are not activated when a new block from a lower memory level is inserted, as well as when a clean or an uncompressed dirty block is read for delivery to the higher cache level.
Hence, no performance overhead is incurred in this regard.
When a block is written back to LLC, the compressor and strong-ECC encoder units are activated, but they do not belong to the critical path.
On eviction of a compressed dirty block from LLC, the decompressor and strong-ECC decoder units are activated, but again, they are not in the critical path and their operations are carried out in the background.
The only source of performance overhead is reading a compressed dirty block when requested by the higher cache level.
As shown in Fig. \ref{fig-ipc}, the proposed scheme reduces the IPC by an average of 0.3\%, normalized to SEC-DED.

Regarding hardware complexity and area overhead, the proposed scheme necessitates the addition of a single bit to the tag of each block to distinguish between compressed and uncompressed blocks.
The storage overhead of this bit is less than 0.2\% for 64-byte cache blocks.
The area and energy consumption of the (de)compressor unit and the strong-ECC encoder/decoder are less than 1\% of those of LLC.
Thus, the proposed scheme delivers a noteworthy improvement in reliability at a negligible cost overhead.


\vspace{-10pt}
\section{Conclusion}\vspace{-5pt}
Emerging racetrack memory (RTM) technology is a promising alternative to SRAMs in LLCs due to its high cell density and comparable access time.
However, the reliability of RTM-based LLCs is compromised by various sources of errors originated from shift operation as well as stochastic switching behaviour of MTJ.
The error correction capability of conventional ECCs is insufficient for highly error-prone RTM-based LLC.
This paper proposes a method of protecting cache blocks by strong ECCs without imposing extra storage for check bits.
This low-cost scheme takes advantage of value locality to compress data blocks and exploits the freed part of cache blocks for storage of strong ECC.
The proposed scheme enhances the MTTF of the cache by 11.3x, on average, compared with the conventional SEC-DED scheme.
The hardware overhead and performance degradation of this scheme is less than 1\%.

\bibliographystyle{IEEEtran}
\bibliography{references-squeez}

\end{document}